\documentclass[11pt]{article}

\usepackage{epsfig}
\usepackage{graphics}
\usepackage{amssymb}

\newcommand{\R}{{\bf R}}
\newcommand{\T}{{\bf T}}

\newcommand{\N}{{\bf N}}
\newcommand{\Z}{{\bf Z}}
\newcommand{\vctr}[2]{\left(\begin{array}{c} {#1} \\ {#2} 
\end{array}\right)}

\newcommand{\mat}[4]{\left(\begin{array}{cc} {#1} & {#2} \\ {#3} & {#4}
\end{array}\right)}

\newtheorem{theorem}{Theorem}

\newtheorem{proposition}{Proposition}
\newtheorem{corollary}{Corollary}
\newtheorem{lemma}{Lemma}
\newtheorem{remark}{Remark}
\newcommand{\proof}{\noindent {\bf Proof} \hspace{0.2in}}
\newcommand{\qed}{\hfill \mbox{\raggedright \rule{.07in}{.1in}}\\
\vspace{0.05in}
}

\begin{document}

\title{Invariant sets for discontinuous parabolic area-preserving
torus maps}

\author{Peter Ashwin and Xin-Chu Fu\\
Department of Mathematics and Statistics\\
University of Surrey\\
Guildford GU2 5XH, UK
\and
Takashi Nishikawa \\
Institute for Plasma Research, University of Maryland\\
College Park, MD 20742, USA
\and
Karol {\.Z}yczkowski\\
 Instytut Fizyki im. Smoluchowskiego, Uniwersytet Jagiello{\'n}ski\\
ul. Reymonta 4, 30-059 Krak{\'o}w, Poland\\
   and\\
Centrum Fizyki Teoretycznej PAN,\\
 Al. Lotnik{\'o}w 32/46, 02-668 Warszawa, Poland\\
}

\maketitle

\begin{abstract}
We analyze a class of piecewise linear parabolic maps on the torus,
namely those obtained by considering a linear map with double 
eigenvalue one and taking modulo one in each component.
We show that within this two parameter family of maps,
the set of noninvertible maps is open and dense. For cases
where the entries in the matrix are rational we show that the maximal
invariant set has positive Lebesgue measure and we give bounds on the
measure. For several examples we find expressions for the measure 
of the invariant set but we leave open the question as to whether there 
are parameters for which this measure is zero.
\end{abstract}

\centerline{{\bf Revised version}}

\vspace{1.2cm}

\centerline{ PACS: {\it 05.45.+b}}
\vspace{0.8cm}


\section{Introduction}

We consider a class of maps of the torus $X=[0,1]^2$ of the form
\begin{equation}\label{eqmap}
\begin{array}{l}
x'=ax+by ~(\bmod ~1)\\
y'=cx+dy ~(\bmod ~1)
\end{array}
\end{equation}
where $(x,y)\in[0,1]^2$. This can be thought of as a map $f=g\circ M$ where
$$
M=\mat{a}{b}{c}{d}
$$
(we also write $M=(a,b;c,d)$ for convenience)
and $g(x)=x-\lfloor x\rfloor$ is a map that takes 
modulo $1$ in each component. Although such
maps are linear except at the discontinuity induced by the map $g$,
their dynamical behaviour can be quite complicated. Depending on
the eigenvalues $\lambda_{1,2}$ of $M$ we refer to the map as 
elliptic ($\lambda_1=\overline{\lambda}_2\neq \lambda_1$), 
hyperbolic ($\lambda_1>\lambda_2$)
or parabolic ($\lambda_1=\lambda_2$). In this paper we focus on the 
area-preserving parabolic case
with determinant  $ad-bc=1$ and trace $a+d=2$. Such maps arise naturally
on examining linear maps with a periodic overflow.
In particular, suppose that one would like to iterate a
matrix using a digital representation with
very small discretization error but a finite range which we set to be
$(0,1)$. If the calculation overflows such that the fractional part remains,
we will get the map (\ref{eqmap}) (compare, for example, with \cite{ACP97}).

In this case $f$ is a piecewise continuous map of $\T^2=\R^2/\Z^2$ to
itself that is area-preserving and such that the linearization is
at almost every point in $\T^2$ is $M$. This matrix has two eigenvalues
equal to one. The map is continuous everywhere on the torus except
on a one-dimensional discontinuity $D$ in $\T^2$. The behaviour of $f$ on
$D$ does not affect a full measure set of $X$.

Parabolic area-preserving maps are not typical in the set of
almost-everywhere linear maps on the torus \cite{Be}. However, 
their dynamical
properties are of a particular interest, since such maps can be
considered as an interpolating case between
the hyperbolic maps ($|t|>2$) and the elliptic  maps ($|t|<2$)
in the area-preserving case. They are in some sense generalisations of 
interval exchange maps \cite{Ke75,Ve78,Ka80},
piecewise rotations \cite{Go2,ACP97} and interval translation maps
\cite{Bos&Kor95,Sch&Tro97} to two dimensions. 
In fact, in Section~\ref{secmaximal} and
the subsequent sections, our results use the fact that
for rational coefficients of $M$
the map (\ref{eqmap}) may be decomposed into a one-parameter family of 1
dimensional interval translation maps.

Hyperbolic area-preserving maps are 
characterized by a positive Liapunov exponents,
and are often studied as model chaotic area-preserving dynamical 
systems (see eg.
\cite{Be,Ott94} and the Baker's transformation). Elliptic area-preserving
maps correspond (in an appropriate eigenbasis) to rigid rotation,
where the presence of a
discontinuity caused by $g$ will lead to very complicated dynamics 
\cite{As96,As97,Go1,ACP97}.
The elliptic--hyperbolic transition for the linear maps on the torus was 
studied by Amadasi and Casartelli \cite{AC91}, while  certain properties of 
linear parabolic maps were analyzed in \cite{Zyc98}.
In particular, a generic parabolic map displays some sort of sensitive
dependence on initial conditions. However, this property is not related to
the Lyapunov exponents (which are always zero for parabolic
maps), but is due to the discontinuity of the map.

\subsection{Summary of main results}

We parametrize the set of parabolic area-preserving maps by $A$ and 
$\alpha$, namely
\begin{equation}\label{eqmapAalpha}
\begin{array}{l}
\vctr{x'}{y'}=f\vctr{x}{y}=
\vctr{(1+A)x+\frac{A}{\alpha}y ~(\bmod 1)}{-\alpha Ax+(1-A)y~(\bmod 1)}
\end{array}.
\end{equation}
Section~\ref{secinvert} 
characterises (in Theorem~\ref{thmnoninvert}) the open dense set of
$(A,\alpha)$ such that the mapping (\ref{eqmapAalpha}) is not invertible
and discuss some properties of the invertible maps.
Section~\ref{secmaximal} examines properties of the 
{\em maximal invariant set} \cite{Zyc98}
$$
X^+=\bigcap_{n=0}^{\infty} f^n(X)
$$
which by the previous result is strictly smaller than $X$ for most
$(A,\alpha)$. Note that any $A\subset X$ with $f(A)=A$ will have
$A\subset X^+$ and so in this sense the set is maximal. 

We say a map (\ref{eqmapAalpha}) is {\em semirational} if $\alpha$
is rational. It is {\em rational} if both $A$ and $\alpha$ are
rational. If $f$ is not semirational we say it is {\em
irrational}. Note that the map is rational if and only if $a$, $b$, $c$ 
and $d$ are rational.

We investigate the two dimensional Lebesgue measure $\ell(\cdot)$ of this
set, showing in Theorem~\ref{thmposmeas} that for
rational parabolic maps we have $\ell(X^+)>0$. We give explicit 
lower and upper bounds for $\ell(X^+)$ depending on $A$ and $\alpha$.

We discuss two particular examples of rational maps where we can
compute $\ell(X^+)$, namely
with $A=\frac{1}{2}$ and $\alpha=1$ or $\frac{1}{2}$.
In the former case we show that $\ell(X^+)=\frac{1}{2}$ and in the
latter case we show that
$\ell(X^+)=\frac{1}{2}-\frac{\sqrt{3}\pi}{72}-\frac{1}{8}\ln 3\sim
0.28710$. In some special cases, e.g., $-1\leq \alpha<0,~ |A| \le 1 $
and $ A = 1, ~ 0< \alpha < 1$ , we can compute $X^+$, and thus we 
can obtain exact values of $\ell(X^+)$.
Other than these cases we have not been able to compute
exact values of $\ell(X^+)$ other than by numerical approximation, even for
rational $A$ and $\alpha$. We suspect that for many $\alpha>0$ and $A>0$ 
(both irrational) then $\ell(X^+)=0$, whereas $\ell(X^+)>0$ for
$\alpha$ rational. Section~\ref{secnonlinexample} briefly discusses 
examples of nonlinear parabolic maps, and we conclude with some remarks
on open problems in Section~\ref{secdiscuss}.

\section{Invertibility of linear maps on the torus}
\label{secinvert}

Consider a matrix $M$ corresponding to a parabolic map on the torus
(\ref{eqmap}). Apart from the special {\em horocyclic} \cite{CFS82}
cases
\begin{equation}
\mat{1}{B}{0}{1}~~~~\mat{1}{0}{B}{1},
\label{horo}
\end{equation}
any such matrix can be written in the form
\begin{equation}
M=M_{A,\alpha}=\mat{1+A}{A/\alpha}{-\alpha A}{1-A},
\label{mapM}
\end{equation}
where $A$ and $\alpha$ are real parameters.
Observe that the cases
$\alpha, A \rightarrow 0$ (with $A/\alpha$ held constant) and
$\alpha\rightarrow\infty$, $A\rightarrow 0$ (with $\alpha A$ held constant)
correspond to the cases (\ref{horo}). Observe that
$$
M_{A,\alpha}^{-1}=M_{-A,\alpha},
$$
and if $S(x,y)=(y,x)$ is transposition then $S^2=I$ and
$$
SM_{A,\alpha}=M_{-A,1/\alpha} S.
$$
Moreover, in the absence of the rounding $g$ we have
$$
M_{A,\alpha}^p=M_{pA,\alpha}
$$
for any $p\in\Z$.

\subsection{Noninvertibility of generic parabolic maps}

The next lemma gives necessary and sufficient conditions that (\ref{eqmap})
is invertible (ignoring points that land on the discontinuity). 
We say the map is invertible if there is a
full measure subset on which it is invertible. Since the mapping
is a composition of an invertible linear map and a discontinuous
map that maps open sets to open sets, both of which
preserve Lebesgue measure, a map is non-invertible
if and only if there is an open set of $(x,y)$ that has
two or more preimages. Let $a,b,c,d$ be defined as previously.

\begin{lemma}\label{lem1}
Any map (\ref{eqmap}) with $ad-bc=1$ will be noninvertible on an open set
if and only if there are $(K,L)\in\Z^2\setminus(0,0)$ integers such that
$$
|Kc-La|<1~\mbox{ and }~|Kd-Lb|<1.
$$
\end{lemma}

\proof
Suppose that we have $(x,y)$ and $(u,v)\in[0,1)^2$ such that $(x,y)\neq (u,v)$
but $f(x,y)=f(u,v)$. Then there is $(K,L)\in \Z^2\setminus (0,0)$
such that $ax+by=au+bv+K$ and $cx+dy=cu+dv+L$. These hold if and only
if
$$
a(x-u)+b(y-v)=K~\mbox{ and }~c(x-u)+d(y-v)=L.
$$
Using the fact that $a,b,c,d$ are non-zero and $ad-bc=1$ this implies that
this holds if and only if
$$
x-u=Kc-La~\mbox{ and }~y-v=Kd-Lb.
$$
Thus the mapping is many-to-one if and only if there are $K$ and $L$
such that
$$
|Kc-La|<1~\mbox{ and }~|Kd-Lb|<1
$$
and the result follows.
\qed

\begin{remark}
It is quite possible that there are simultaneously many solutions to
the inequalities of Lemma~\ref{lem1}; for example, near $A=-1$ and
$\alpha$ large one can find arbitrarily large numbers of integers
$(K,L)$ satisfying both inequalities.
\end{remark}

Using the previous Lemma we obtain the main result of this section.

\begin{theorem}\label{thmnoninvert}
The only invertible parabolic maps $f$ have $M$ equal to one of
$$
\mat{k}{-(k-1)^2/l}{l}{2-k},~~
\mat{2+l}{-k}{(l+1)^2/k}{-l},~~
\mat{1}{b}{0}{1}~~\mbox{ or }~~
\mat{1}{0}{b}{1},
$$
where $(k,l)$ are integers and $b$ is real.
\end{theorem}

\proof
The square $V_\epsilon=(-1-\epsilon,1+\epsilon)^2$ is
a convex region symmetrical about the origin with area $4(1+\epsilon)^2$. A
variant of Minkowski's theorem \cite[Thm 447]{Har&Wri79}
says that for any $\epsilon>0$ there will be a point on the lattice
$\Lambda=\{(Kc-La,Kd-Lb)~:~(K,L)\in\Z^2\}$ other than zero inside $V_\epsilon$,
if the lattice has determinant $\Delta=ad-bc\geq 1$. Taking the limit 
$\epsilon\rightarrow 0$ and using compactness we see that 
there must be a point $(\kappa,\lambda)\in \Lambda$ other than zero
in $\overline{V_0}$. 

If there is a lattice point $(\kappa,\lambda)$ in the interior of
$V_0$, we are done. If not, assume without loss of generality that 
$(\kappa,\lambda)=(1,y)$ with $|y|\leq 1$.
Thus exactly one point in the line segment from (but not including)
$(-1,1-y)$ to $(0,1)$ that is in the lattice (one can generate the 
same lattice by adding multiples of $(1,y)$ to any lattice point
generating the lattice with $(1,y)$). Therefore there
will be a point in the lattice in the interior of $V_0$ 
unless this other point is $(0,1)$. Therefore (considering the other case by 
interchanging $x$ and $y$) there will be integers $K$ and $L$ such 
that either
$$
(Kc-La=1~~\mbox{ and}~~Kd-Lb=0)~~~\mbox{ or }~~~
(Kc-La=0~~\mbox{and}~~Kd-Lb=1).
$$
Thus, in the invertible case the matrix $M$ is determined by which of 
these cases occurs. One can easily solve to show that either of these
cases can occur. If $k=1$ or $l=-1$ then the other (horocyclic)
cases occur.
\qed

It follows from this result that most parabolic maps are
noninvertible in the following sense:

\begin{corollary}
The set of noninvertible maps is open and dense within the set of
parabolic maps.
\end{corollary}

Because the $(\bmod ~1)$ map  $g$ commutes with integer matrix and $g^2 = g$, 
all conjugations of parabolic maps with integer coefficients by automorphisms 
of the torus  $\T^2$  remain integer parabolic maps. For an integer parabolic 
map with rational $\alpha = r/s$, where $(r, ~s) = 1$, since there 
exist $m,~n \in \Z$ such that $mr+ns=1$, there exists an automorphism whose 
$GL(2,\Z)$ matrix has bottom row $(r,~s)$. Conjugation by this automorphism 
gives a parabolic map whose matrix has bottom row $(0,~*)$. 
Since conjugation 
preserves integrality, trace and determinant, the resulting matrix must be of 
the form $(1,~B;~0,~1)$ for some $B$. Therefore we conclude that all parabolic 
maps with integer coefficients can be reduced to one of the horocyclic cases.

\section{Maximal invariant sets}
\label{secmaximal}

By Poincar\'{e} recurrence the maximal invariant set
$X^+=X$ (up to a set of zero measure) if and only if $f$ is 
invertible at almost every point. In fact the maximal invariant 
set $X^+$ is an upper semicontinuous function of the
system parameters $A,\alpha$ in the Hausdorff metric.
However, the measure (and dimension) of $X^+$ can (and does) 
change discontinously with parameters; see 
Proposition~\ref{propsimplecases}.

We now consider the structure of $X^+$ for the semirational
case (ie $\alpha$ rational). Consider the family 
of lines $L_B$ defined by
$$
y = B- \alpha x
$$
with $B$ a fixed real number.
projected onto the torus by taking modulo $1$ in $x$ and $y$. 
This can be thought of as the set
$$
L_B=\{ (x,y)~:~ y=B+K-\alpha (x+L)~~\mbox{with}~~(K,L)\in\Z^2\}\cap X.
$$
Note that if 
$$
y=B+K-\alpha (x+L)
$$
with $L$ and $K$ integers such that $x,y$ in $[0,1]$  then
$y'=-\alpha A x+ (1-A)(B+K-\alpha (x+L) +N$ and
$x'=(1+A) x + A/\alpha (B+K-\alpha (x+L))+M$ with $M,N$ integers.
Rearranging this we have
$$
y'= B+(K+M+N)-\alpha(x+L)
$$
and so the family of lines $L_B$ is invariant under the map for any given $B$.
We define the maximal invariant set within $L_B$ as
$$
X_B^+= X^+ \cap L_B.
$$

\subsection{Semirational and rational parabolic maps}\label{secrational}

For semirational maps ($\alpha=r/s>0$) and for any given value of $B$ 
the set $L_B$ consists of $r+s$ (or exceptionally $r+s-1$ if 
it contains the origin)	intervals that are parallel to the 
eigenvector $v=(1,-\alpha)$ of the 
matrix $M$. We can parametrise any particular $L_B$ by $\theta\in[0,s)$.
We will consider $\alpha>0$ from here on.
More precisely we re-parametrize $X$ by
\begin{equation}
\begin{array}{l}
x = \theta - [\theta]\\
y = B- \alpha\theta -[B-\alpha\theta]
\end{array}
\end{equation}
For $\theta\in[0,s)$ and $B\in [\frac{s-1}{s},1)$ the mapping 
$(x,y) \leftrightarrow (\theta,B)$ is
one-to-one and has unit Jacobian everywhere.

Since each $L_B$ is invariant the maximal invariant set must nontrivially
intersect $L_B$ and so no single trajectory is dense in $X^+$. In this case
we need to approximate $X^+$ using a distribution of trajectories
on a dense set of lines.

The relation between \( \theta  \) and \( x \) is 
\( x=\theta -[\theta ]. \)
The variable \( y \) is related to \( \theta  \) and \( B
\) by the 
relation \( y=B-\alpha \theta -[B-\alpha \theta ] \).
So the point $(\theta ,B)$ maps to
$$
(\theta',B')=
(\theta +AB/\alpha -A[\theta ]-A/\alpha [B-\alpha \theta ],B) 
$$
since
\begin{eqnarray*}
\vctr{\theta'}{v} &=&
\left( \begin{array}{cc}
1+A & A/\alpha \\
-A\alpha  & 1-A
\end{array}\right) \left( \begin{array}{c}
\theta -[\theta ]\\
B-\alpha \theta -[B-\alpha \theta ]
\end{array}\right) +\left( \begin{array}{c}
[\theta ]\\
\left[ B-\alpha \theta \right] 
\end{array}\right) \\
&=& \left( \begin{array}{c}
\theta +AB/\alpha -A[\theta ]-A/\alpha [B-\alpha \theta ]\\
-\alpha \theta +B-AB+A\alpha [\theta ]+A[B-\alpha \theta ]
\end{array}\right). 
\end{eqnarray*}
We have added integers
\( [\theta ] \) and \( [B-\alpha \theta ] \) so that
$\theta'\in[0,s)$ and $v=B-\alpha\theta$ (see eg \cite{Zyc98}).

Thus the action of the map (\ref{eqmap}) on 
the coordinate $\theta$ is simply
\begin{equation}
\theta'=T_B(\theta)=\left(\theta+\frac{s}{r}AB-A[\theta]-A\frac{s}{r}
[B-\frac{r}{s}\theta] \right) ~ (\bmod s).
\label{map1d}
\end{equation}
Because the variable $B\in [\frac{1-s}{s},1)$ is invariant during the
dynamics it is treated as a parameter.  Again, the slope of these
1D maps is equal to one implying that all Liapunov exponents are zero.

In the case of a rational map, ie where 
both $A=p/q$ and $\alpha=r/s$ are rational with
$(r,s)=(p,q)=1$ we can understand a lot about the mapping $T$ using the 
following factor map. Let 
$$
\pi(\theta)= \theta-\frac{1}{qr}[qr\theta]
$$
be a projection of $I=[0,s)$ onto $J=[0,\frac{1}{qr})$: this maps
$qrs$ points onto one point.

\begin{lemma}
If $A=\frac{p}{q}$ and $\alpha=\frac{r}{s}$ then the diagram
$$
\begin{array}{rcl}
I & \stackrel{T_B}{\rightarrow} & I \\
\pi \downarrow & & \downarrow \pi  \\
J & \stackrel{S_B}{\rightarrow} & J 
\end{array}
$$
commutes where 
$$
S_B(\psi)=\psi+ \frac{sp}{rq}B ~ (\bmod \frac{1}{qr})
$$
is a rotation. In other words $T_B$ has a factor that is a
rotation.
\end{lemma}

\proof
For these assumptions we can write (\ref{map1d}) as
\begin{equation}
\theta'=\left( \theta+\frac{sp}{rq}B-\frac{p}{q}[\theta]
-\frac{ps}{qr}[B-\frac{r}{s}\theta]\right) ~(\bmod s).
\label{map1drational}
\end{equation}
Defining $\psi=\pi(\theta)$ and $\psi'=\pi(\theta')$, note that 
$$
\psi'=\psi+ C ~(\bmod \frac{1}{qr})
$$
where $C=\pi(AB/\alpha)=\pi(\frac{sp}{rq}B)$. 
\qed

Since $S_B$ is invertible, $\cap_{n=0}^{\infty} S^n(J)=J$.
The map $\pi$ maps $M=qrs$ points to one point and so
$\pi^{-1}$ must be understood as a set-valued function. We have
$$
\pi\pi^{-1}(\psi)=\psi,~~ \pi T (\theta) = S \pi(\theta)
$$
for all $\psi\in J$ and $\theta\in I$, while for
any set $K\subset I$
$$
\pi^{-1}\pi K \supseteq K.
$$
Let
\begin{eqnarray*}
N(\psi) &=& \{ \theta\in X^+_B~:~ \pi(\theta)=\psi\} = 
\pi^{-1}(\psi) \cap X_B^+
\end{eqnarray*}
be the set of $\theta\in X_B^+$ with $\pi(\theta)=\psi$ 
(we suppress the dependence of $S$ and $T$ on $B$ for the next result).

\begin{lemma}
For any $\psi\in J$, the set $N$ satsifies
$$
N( S(\psi)) = T ( N(\psi) ).
$$
\end{lemma}

\proof
To show that $T(N(\psi))\subseteq N(S(\psi))$, note that
\begin{eqnarray*}
T(N(\psi)) & \subseteq & \pi^{-1} \pi T(N(\psi)) = \pi^{-1} S \pi (N(\psi))\\
& = & \pi^{-1} S \pi (\pi^{-1} (\psi) \cap X_B^+) \\
& \subseteq  & \pi^{-1} S(\psi)) \cap X_B^+ \\
& =  & N(S(\psi)).
\end{eqnarray*}
For the other direction, 
suppose that $\theta\in N(S(\psi))$ and so $\pi(\theta)=S(\psi)$ and
$\theta\in \cap_{n\geq 0} T^n(I)$.  In particular, $\theta=T(\theta')$ 
for some $\theta'$. Now $\pi(\theta) = \pi T(\theta')= S \pi(\theta')$ 
and so $S(\psi)= S \pi(\theta')$. Invertibility of $S$ gives 
$\psi=\pi(\theta')$ and so $\theta'\in N(\psi)$. Hence
$$
T(N(\psi)) \supseteq  N(S(\psi))
$$
and we have the result.
\qed

The previous Lemma relies crucially on the fact that $S$ is invertible.
The next result implies that the number of points in $N$ is a constant almost 
everywhere. Let
$$
\hat{N}=|N(\psi)|
$$
be the cardinality of $N(\psi)$.

\begin{lemma}
If $B$ is irrational then $\hat{N}$ is constant for a set of $\psi$
with full Lebesgue measure.
\end{lemma}

\proof
It is a standard result that $S_B$ is ergodic for Lebesgue measure
if and only if $B$ is irrational. If we look at the set of $\psi$
that give a certain value of $N(\psi)$ this is invariant and therefore
must have Lebesgue measure $0$ or $1$.
\qed

One consequence of this is that the measure of $X_B^+$ can only
take a finite number of values. We write $\hat{N}(B)$ to show the
dependence on $B$ explicitly.

\begin{theorem}\label{thmxb+meas}
Suppose that $A=p/q$ and $\alpha=r/s$. For Lebesgue almost all $B$ we have
$$
\ell(X_B^+) = \frac{\hat{N}(B)}{qr}
$$
where $\hat{N}(B)$ is an integer and $1\leq \hat{N}(B)\leq qrs$. If 
there is an interval
on which $T$ has $L$ preimages, then
$$
\hat{N}(B)\leq qrs+1-L.
$$
\end{theorem}

\proof
As preimages of $\pi^{-1}(J)$ are disjoint and all have length $\frac{1}{qr}$
we compute
$$
\ell(X_B^+) = \sum_{k=1}^{\hat{N}(B)} \frac{1}{qr} = \frac{\hat{N}(B)}{qr}.
$$
Note that $\hat{N}(B)$ must take an integer value less than or equal to
$qrs$, being the number of preimages $\pi^{-1}(\psi)$. 

Since $T$ will be invertible on $X_B^+$, if there is an interval with
more than one preimage, only one of these will be in $X_B^+$ and hence
$\hat{N}(B)\leq qrs+1-L$ where $L$ is the number of preimages.
\qed

We can now prove the main result in this section.

\begin{theorem}\label{thmposmeas}
Suppose that $\alpha=r/s$ and $A=p/q$ are both rational, with
$(r,s)=(p,q)=1$.
Then the maximal invariant set $X^+$ has Hausdorff dimension $2$ and
positive Lebesgue measure, more precisely, if there is an
open set
on which $f$ has $K$ preimages then
$$
\frac{1}{qrs}\leq \ell(X^+) \leq 1-\frac{K-1}{qrs}.
$$
\end{theorem}

\proof
Note that under the hypotheses of the theorem,
\begin{eqnarray*}
\ell(X^+) &=& \int_{B\in[1-\frac{1}{s})}\int_{s\in X_B^+}
\,d\theta\,ds
\\
&=& \int_{B\in[1-\frac{1}{s})}\ell(X_B^+)\,ds.
\end{eqnarray*}
Upon integrating the estimates in Theorem~\ref{thmxb+meas} we 
obtain the result.
\qed

We can characterise the dynamics on the lines $L_B$ in the following 
way:

\begin{corollary}
For any rational map,
almost all $B$ and all $x\in L_B$, $\omega(x)$ consists of a set with
positive one-dimension measure in $L_B$. Conversely, for a countable set 
of $B$ and all $x\in L_B$, the trajectory going through $x$ is eventually
periodic.
\end{corollary}

\proof
In the proof of Theorem~\ref{thmposmeas}, the reduced map $S_B$
will be an irrational
rotation for all irrational $B$ and a rational rotation for all rational
$B$. The result follows.
\qed

Density of irrational $B$ simply implies the following result 
for the original map.

\begin{corollary}
For any rational parabolic map, the
preimages of the discontinuity are dense in $X^+$.
\end{corollary}

We do not have a precise analytical expression for $\ell(X^+)$ even
for rational parabolic maps. Numerically one can approximate
the measure and obtain various values of $\ell(X^+)$; see
for example Table~\ref{tablevals}. In certain simple cases
one can obtain exact values of $\ell(X^+)$ by 
constructing the maximal invariant set explicitly.

\begin{proposition}
\label{propsimplecases}
One can compute
$$
\ell(X^+)= \left\{
\begin{array}{ll}
1 & \mbox{ if the map is invertible,}\\
\alpha & \mbox{ if $A=1$ and $0<\alpha\leq 1$,}\\
\alpha^{-1} & \mbox{ if $A=-1$ and $\alpha\geq 1$, and}\\
|A\alpha| & \mbox{ if $0<A\leq 1$ and $-1\leq \alpha <0$}.
\end{array}\right.
$$
\end{proposition}

The invertible case is trivial, while the cases where $A=1$
follow because the map reduces to an invertible map on a strip height $\alpha$.
The appendix gives a constructive proof for the case $0<A\leq 1$,
$-1\leq \alpha<0$. Because $f_{-A,-1}$ is conjugate to  $f_{A,-1}$ by
interchanging $x$ and $y$ we have
$$
\ell(X^+)=\left\{ 
\begin{array}{cc}
|A|~, &  ~~ 0<|A|\le 1~, \alpha=-1\\
 1~,  &  ~~ A=0 ~, \alpha=-1.
\end{array}\right.
$$ 
Note in particular that there is a discontinuity 
in $\ell(X^+)$ at $(\alpha,A)=(1,0)$.

The following examples (especially
that in Section~\ref{secegii}) show that a general expression, if it
can be obtained, is likely to be non-trivial.

\begin{table}
$$
\begin{array}{l|llll}
A & \alpha=2/3 & 1/2 & 1/3 & 1/5 \\
\hline
1/4 &  0.261  & 0.337  & 0.251  & 0.196 \\
1/3 &  0.235  & 0.326  & 0.265  & 0.199 \\
2/3 &  0.215  & 0.342  & 0.257  & 0.149 \\
5/4 &  0.628  & 0.626  & 0.419  & 0.252 \\
3/2 &  0.338  & 0.562  & 0.505  & 0.303 
\end{array}
$$
\caption{\label{tablevals}
Numerically obtained approximate values of $\ell(X^+)$ for a number
of rational parabolic maps. }
\end{table}

\subsection{Example I}

\begin{figure}

\begin{center}
\epsfig{file=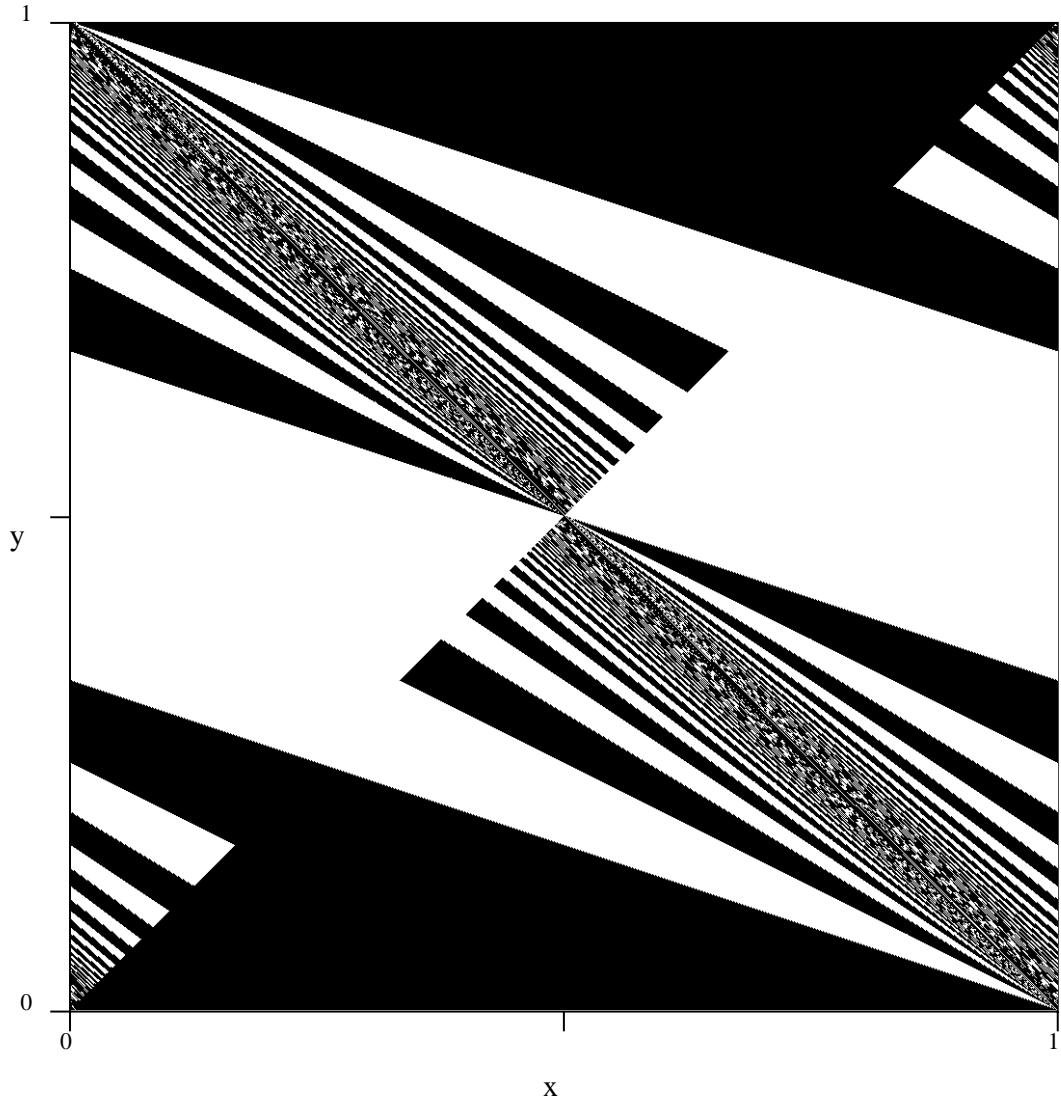,width=14cm}
\end{center}

\caption{\label{figa1_2alp1}
The black region shows the
maximal invariant set $X^+$ for the map ({\protect \ref{eqeg0}}) where
$A=1/2$ and $\alpha=1$. Observe the symmetry of the black and white regions
indicating that $\ell(X^+)=\frac{1}{2}$.}
\end{figure}

The simplest nontrivial rational map is defined by $A=1/2$ and $\alpha=1$.
In this case, we can approximate the maximal invariant set
numerically as in Figure~\ref{figa1_2alp1}. The map in this case is
\begin{equation}\label{eqeg0}
\begin{array}{l}
x'=\frac{3}{2}x+\frac{1}{2}y ~(\bmod 1)\\
y'=-\frac{1}{2}x+\frac{1}{2}y~(\bmod 1)
\end{array}
\end{equation}
where $(x,y)\in X=[0,1]^2$. In this case, as noted in \cite{Zyc98},
there appears to exist a symmetry between $X^+$ and its
complement. In fact, we can use the results
in Section~\ref{secrational} to get the result directly.

\begin{proposition}
For the map (\ref{eqeg0}) with $A=\frac{1}{2}$ and $\alpha=1$,
we have $\ell(X^+)=\frac{1}{2}$.
\end{proposition}

\proof
Note that $r=s=p=1$ and $q=2$. Moreover, on $L_B$
the map (\ref{map1d}) can be written
$$
\theta'=h(\theta)=\theta+\frac{B}{2} - \frac{1}{2}[B-\theta]~(\bmod 1)
$$
and this map is two-to-one on an open set of $\theta$ for almost
all $B$. Therefore, for almost all $B$, the upper and
lower bounds of Theorem~\ref{thmxb+meas} agree and
$$
\ell(X_B^+)=\frac{1}{2}
$$
for almost all $B$. This implies that $\ell(X^+)=\frac{1}{2}$.
\qed

\subsection{Example II}\label{secegii}

We now consider the case $A=\alpha=1/2$; this shows that
$\ell(X^+)$ may be irrational for a rational map. In this case
the map is
\begin{equation}\label{eqeg1}
\begin{array}{l}
x'=\frac{3}{2}x+y ~(\bmod 1)\\
y'=-\frac{1}{4}x+\frac{1}{2}y~(\bmod 1)
\end{array}
\end{equation}
On the invariant lines $L_B$ with
$B\in[\frac{1}{2},1)$ (\ref{eqeg1}) reduces to (\ref{map1d}) with
$\theta\in[0,2)$ to
\begin{equation}
\theta' = T_B(\theta)=\left(\theta+B-\frac{1}{2}[\theta]-
[B-2\theta] \right) ~ (\bmod 2).
\label{pmap}
\end{equation}

\begin{figure}
\setlength{\unitlength}{1in}
\begin{picture}(6,6)(0,0)
\put(0.5,0.53){\resizebox{5in}{4.96in}{\includegraphics{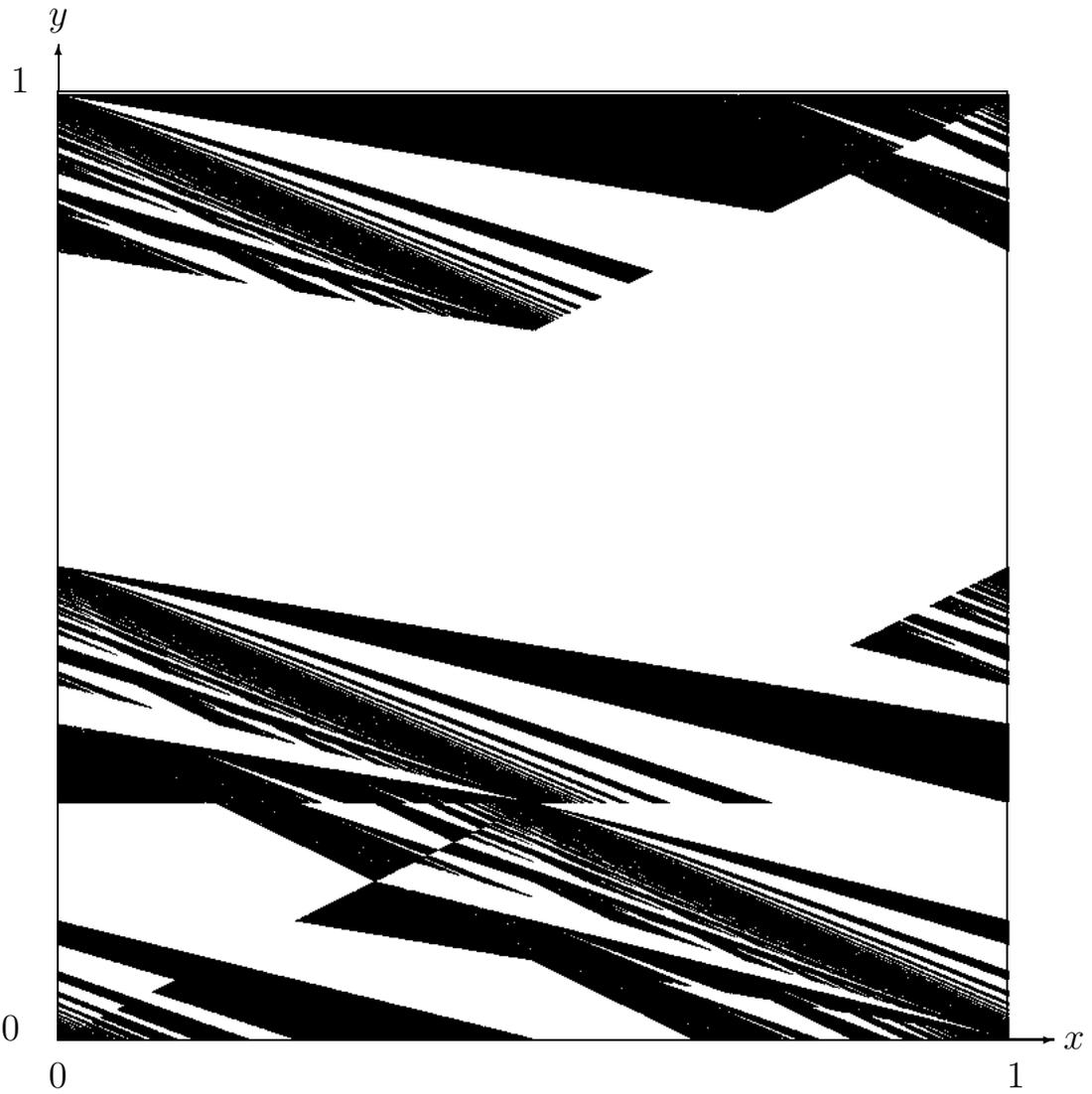}}}
\put(0.2,0.5){\Large{0}}
\put(0.45,0.25){\Large{0}}
\put(0.25,5.5){\Large{1}}
\put(5.5,0.25){\Large{1}}
\put(0.5,0.5){\framebox(5,5)}
\put(0.5,5.5){\vector(0,1){0.25}}
\put(5.5,0.5){\vector(1,0){0.25}}
\put(0.45,5.85){\Large{$y$}}
\put(5.8,0.45){\Large{$x$}}
\end{picture}
\label{fig:fig1}
\caption{The black regions shows the maximal invariant
set $X^+$ of the rational map (9) where $A = \alpha
= \frac{1}{2}$.
This has measure corresponding to approximately 28.7\% of the
unit square.
}
\end{figure}

The maximal invariant set $X^+$ for the map
(\ref{eqeg1}) is shown in Fig. 2. The results in Theorem~\ref{thmposmeas}
imply that
$$
\frac{1}{4}\leq \ell(X^+)\leq \frac{3}{4}
$$
but this is clearly not very precise. 

We have obtained a numerical estimate by iterating a grid of initial
points distributed uniformly on
the square, dividing the unit square into boxes and counting
the occupied cells. To avoid the transient effects some 
number (say, the first hundred)
of images are not marked on the graph.
Applying this method we obtained $V\sim 0.287$, however the 
precision of this result is limited by a highly complicated
structure of some fragments of the invariant set and a very slow
convergence of transients.  

\begin{figure}
\setlength{\unitlength}{1in}
\begin{picture}(6,6)(0,0)
\put(0.5,0.53){\resizebox{5in}{4.96in}{\includegraphics{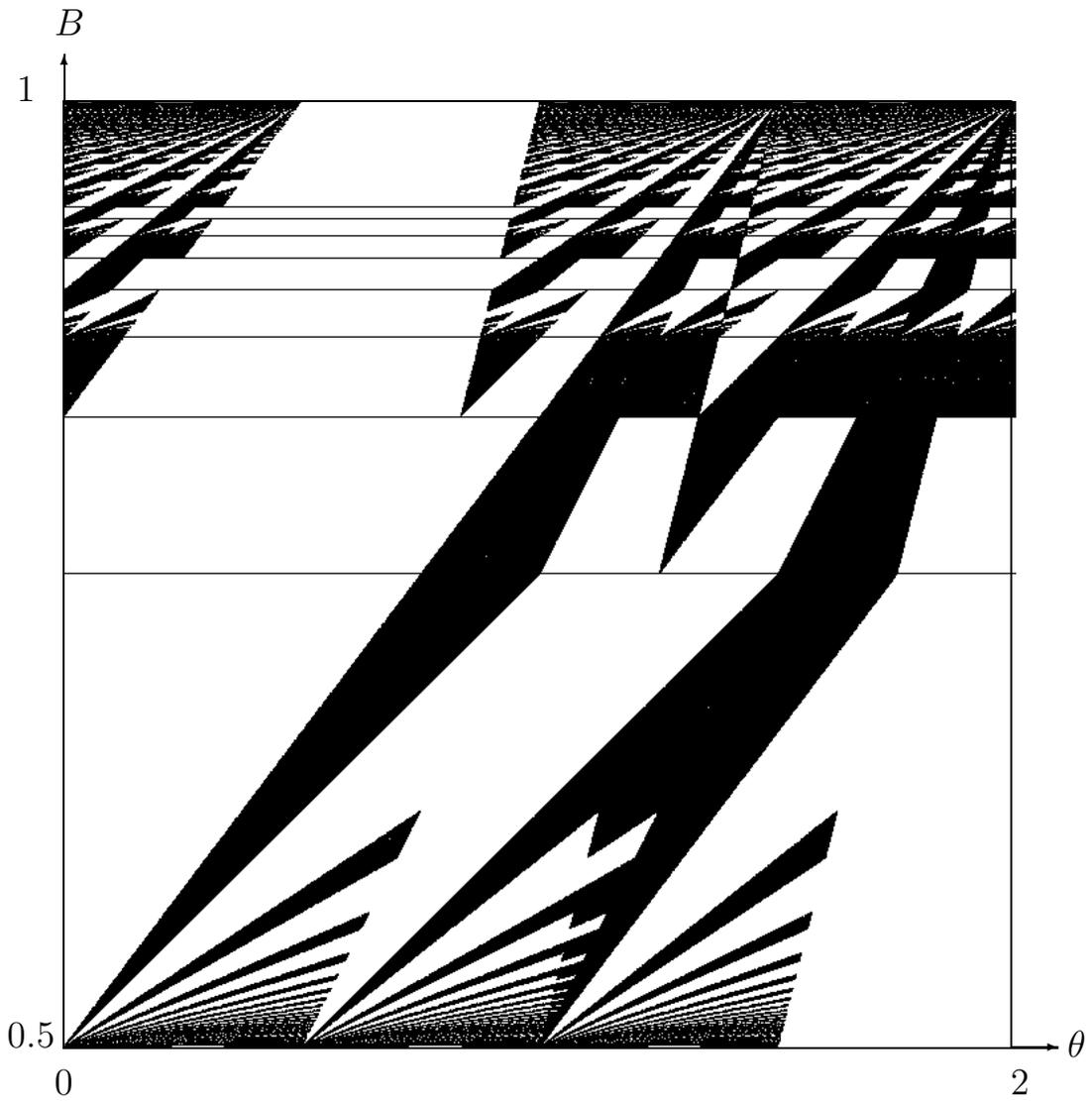}}}
\put(0.2,0.5){\Large{0.5}}
\put(0.45,0.25){\Large{0}}
\put(0.25,5.5){\Large{1}}
\put(5.5,0.25){\Large{2}}
\put(0.5,0.5){\framebox(5,5)}
\put(0.5,5.5){\vector(0,1){0.25}}
\put(5.5,0.5){\vector(1,0){0.25}}
\put(0.45,5.85){\Large{$B$}}
\put(5.8,0.45){\Large{$\theta$}}
\end{picture}
\label{fig:bif}
\caption{Maximal invariant set shown as a bifurcation diagram of 
the map (\ref{pmap}). The
horizontal lines $B = {3 \over 4}, {5 \over 6}, {7 \over 8}, {9 \over
10}, {11 \over 12}, {13 \over 14}, {15 \over 16}$ are superimposed
onto the picture to make the assertion in the
text easier to see.}
\end{figure}

By using the structure noted in Section~\ref{secmaximal} we show
numerical approximations of
$X_B^+$ as $B$ changes in the map (\ref{pmap}) are
shown in Figure.~\ref{fig:bif} and notice that
the bifurcation diagram of $X_B^+$ has a regular structure
$$
\begin{array}{cc}
{1 \over 2} < B < {5 \over 6} & \ell(X^+_B) = {1 \over 2}\\
{5 \over 6} < B < {7 \over 8} & \ell(X^+_B) = 1\\
{7 \over 8} < B < {11 \over 12} & \ell(X^+_B) = {1 \over 2}\\
{11 \over 12} < B < {13 \over 14} & \ell(X^+_B) = 1\\
{13 \over 14} < B < {17 \over 18} & \ell(X^+_B) = {1\over 2}\\
\end{array}
$$
These values agree with the predictions of Theorem~\ref{thmposmeas}
that $\ell(X^+_B)\in \frac{1}{4}\{1,2,3\}$.
More generally for any $n \in \N$ we observe that
$$
\begin{array}{ll}
{6n-1 \over 6n} < B < {6n+1 \over 6n+2} & \ell(X^+_B) = {1}\\
     {\rm else} & \ell(X^+_B) = {1 \over 2}.\\
\end{array}
$$
Therefore we have
\begin{eqnarray*}
\ell(X^+) &=& \int_{1/2}^1 \ell(X_B^+) dB \\
&=& \frac{1}{4}+\frac{1}{2}\sum_{n=1}^\infty
\left({6n+1 \over 6n+2} - {6n-1 \over 6n}\right) \\
&=& {1 \over 4} + {1 \over 12}
\sum_{n=1}^\infty {1 \over 3n^2+n} \\
&=& {1 \over 2} - {\sqrt{3}\pi \over 72} - {1 \over 8} \ln 3 \\
&\approx& 0.28709849.
\end{eqnarray*}
Thus we can obtain a series expression for the measure and in this
case we can sum it explicitly. Unfortunately we have not been able to
extend this method to one that applies generally.

\subsection{Irrational parabolic maps}

We have not yet obtained any significant results for irrational parabolic 
area-preserving maps and, in fact, for the all cases where
we can compute it, the maximal invariant set has positive measure.
Nonetheless, there is numerical evidence (\cite{Zyc98}) that
suggests that 
\begin{itemize}
\item
There are $(\alpha,A)$ for which the Hausdorff dimension of $X^+$ is 
less than 2 (in particular, such that $\ell(X^+)=0$). By results
here, this can only occur in cases where both parameters
are irrational.
\item 
This is common for $A>0$ and $\alpha>0$.
\end{itemize}
Note that not all irrational parabolic maps have zero measure; for example
Proposition~\ref{propsimplecases} shows that there are open
regions in $(A,\alpha)$ with $\alpha<0$ where $\ell(X^+)>0$. For semirational
cases ($\alpha=r/s$) we can reduce to 
maps of the form (\ref{map1d}), however we cannot find factors that are 
rotations if $A$ is not rational. Results of \cite{Bos&Kor95,Sch&Tro97} 
suggest that the Hausdorff dimension of $X^+$ may vary between
$1$ and $2$. The Hausdorff dimension is a rather subtle characteristic of a set.
For example, two sets may be arbitrarily close in the sense
of the Hausdorff metric, but their dimensions may differ by an
arbitrarily large number. Therefore, small changes of the parameter $\alpha$,
which do not change much the structure of the maximal
invariant set (in sense of the Hausdorff metric)
may induce wild fluctuations of its dimension. Note also that the cases 
$\alpha>0$ and $\alpha<0$ seem to be fundamentally different; we do not have
a way of conjugating one with the other.

\section{Nonlinear parabolic maps}
\label{secnonlinexample}

One can also construct nonlinear parabolic maps 
that display similar structure in their maximal invariant sets as
the piecewise linear case.
Consider the two parameter family of the maps on the torus defined by
\begin{equation}
\label{nonlin}
\begin{array}{l}
x'=x+ \beta(e^{\gamma (y-x)} -1)~~(\bmod ~1)\\
y'=y+ \beta(e^{\gamma (y-x)} -1)~~(\bmod ~1)\\
\end{array}
\end{equation}
For any non-zero values of the real parameters $\beta$ and $\gamma$
the Jacobian $\partial(x',y')/\partial(x,y)$ is constant and equal to one.
Moreover, for any point
$(x,y)$ the trace of the map $t$ is equal to $2$, which makes the above
map similar to the linear case (\ref{eqmap}).
Observe that the diagonal of the unit square $y=x$ is
invariant with respect to this map. This map acting on the plane
transforms the unit square into a set confined between four
exponential functions. After folding it back into the unit square some
fragments of it will overlap as (\ref{nonlin}) is typically
noninvertible.

\begin{figure}
\setlength{\unitlength}{1in}
\begin{picture}(6,6)(0,0)
\put(0.5,0.5){\resizebox{5.06in}{5.15in}{\includegraphics{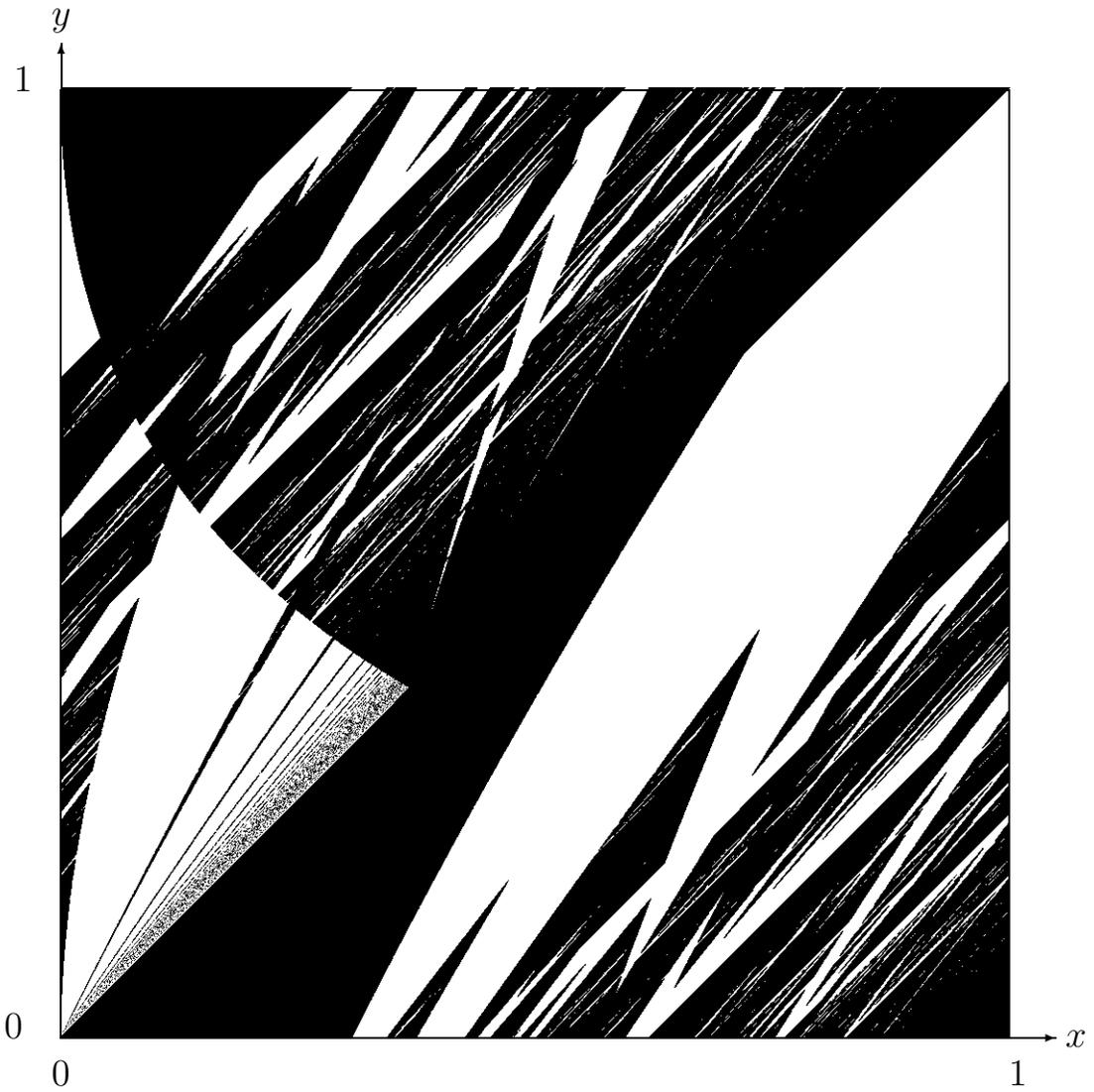}}}
\put(0.2,0.5){\Large{0}}
\put(0.45,0.25){\Large{0}}
\put(0.25,5.5){\Large{1}}
\put(5.5,0.25){\Large{1}}
\put(0.5,0.5){\framebox(5.0,5.0)}
\put(0.5,5.5){\vector(0,1){0.25}}
\put(5.5,0.5){\vector(1,0){0.25}}
\put(0.45,5.85){\Large{$y$}}
\put(5.8,0.45){\Large{$x$}}
\end{picture}
\label{fignonlin}
\caption{The black region shows the maximal invariant
set $X^+$ for the nonlinear parabolic map (\ref{nonlin})
with $\alpha = \gamma = 1$.
}
\end{figure}

Figure~\ref{fignonlin} shows the invariant set of the nonlinear map in
the case $\beta=\gamma=1$. There is numerical
evidence that the maximal invariant set has positive volume in this
case even though the Jacobian changes between being a rational and 
an irrational parabolic map at different points.

\section{Discussion}
\label{secdiscuss}

In this paper we have considered some basic properties of the maps
(\ref{eqmap}) in the parabolic area-preserving regime, and show 
a surprising degree of sensitivity of the dynamical behaviour on 
rationality of the parameters $\alpha$ and $A$. 
Our investigations raise some intriguing 
questions concerning how the Lebesgue measure of the maximal invariant set
varies with parameters. For several examples do we 
have analytical	values of this measure. For general rational
parabolic maps we do have a result that gives upper and lower bounds for
this measure. However, these bounds become weak as one
examines higher denominator rational mapsand do not easily give
insight into the measure for irrational parabolic maps.

It would be very informative to understand the structure of
invariant measures of the map (\ref{eqmap}). Note that $\ell(.)$
(Lebesgue measure) restricted to $X^+$ is invariant under this mapping
but it is not ergodic. In particular, it may be the case that this
restriction is trivial in which case empirical measures
can still be defined; by analogy with the Interval Translation Maps
of \cite{Bos&Kor95,Sch&Tro97} we surmise that there are cases where a
Hausdorff measure restricted to $X^+$ is invariant, in particular in
cases where $\ell(X^+)=0$.

\section*{Acknowledgements}

The research of PA and XF is supported by EPSRC grant
GR/M36335, while K{\.Z} acknowledges the support by a KBN
Grant. PA, TN and K{\.Z} thank the Max-Planck-Institute for the
Physics of Complex Systems, Dresden for providing an opportunity to meet 
and discuss this paper.
We also thank the anonymous referees for their invaluable comments,
in particular to one who pointed out a considerable simplification
of our argument in Section~2.

\section*{Appendix: Proof of Proposition~\ref{propsimplecases}}
\label{secexample3}

\begin{figure}[h]
\begin{center}
\epsfig{file=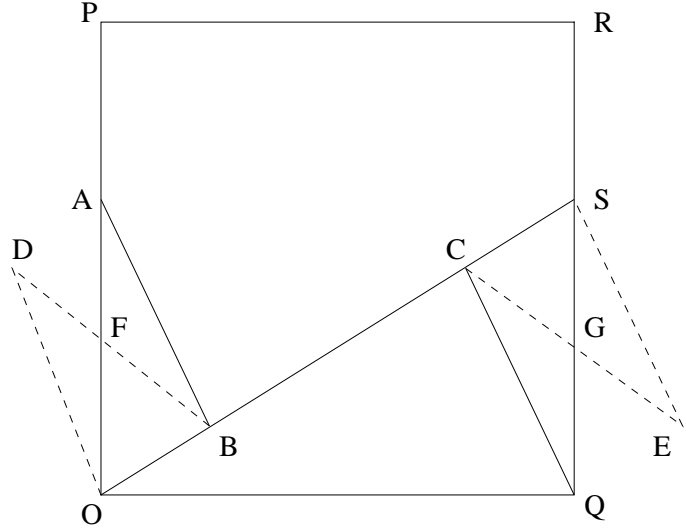, width=9cm}
\end{center}
\caption{ \label{fig_paraconst} Construction of the
maximal invariant set for parabolic maps with $-1\leq\alpha<0$ 
and $0<A\leq1$.}
\end{figure}

We consider the case $-1\leq \alpha<0$ and $0<A<1$;
figure~\ref{fig_paraconst} shows how  the linear map $M_{A, \alpha}$ maps
the square $[0,1]^2$ given by $OPRQ$ into the maximal invariant set 
consisting of the union of the line $OS$ and the triangles 
$OAB$ and $CSQ$, where
$$
A=(0,-\alpha),~~B=(A,-\alpha A),~~C=(1-A,-\alpha(1-A))~~\mbox{ and }~~
S=(1.-\alpha).
$$
To see this is the maximal invariant set, note that everything in $OPRS$ 
decreases in its $y$- component
unless it lands in the triangle $OAB$. Similarly, all points in $OQS$
must increase in $y$-component unless it lands in the triangle $CQS$.
The union of the two triangles is invariant as the dotted images
$ODB$ and $CSQ$ show, as is the line of fixed points $OS$.
Hence in this case we have
$$
\ell(X^+)=-\alpha A= |\alpha A|.
$$
A similar argument (with extra triangles) can be used to show that 
for $0<A \leq 1$ and
$$
-2 < \frac{-2}{1+A} \leq \alpha<-1
$$
the maximal invariant set has measure
$$
\ell(X^+)=-\frac{\alpha A}{2},
$$
but we omit this for conciseness.

\end{document}